\documentclass[smus]{snow2e}
\usepackage{epsfig}
\def\beq{\begin{equation}}
\def\eeq{\end{equation}}

\def\beqn{\begin{eqnarray}}
\def\eeqn{\end{eqnarray}}
\begin{document}
\title{Probing Quartic Couplings Through
Three\\
Gauge Boson Production at an $e^+e^-$ Linear Collider
\thanks{The work of  S.D.   supported by 
DOE contract number DU-AC02-76CH00016.  
The work of  G.V. supported in part by the DOE OJI
program under contract number DE-FG02-92ER40730. The
work of O.Y. and A. L. supported in part by RFBR under grant
96-02-18216.  The work of A.L. has been made possible by
a fellowship of INTAS grant 93-2492-ext and is carried
out within the research program of ICFPM. }}
\author{S. Dawson$^a$, A. Likhoded$^b$, G. Valencia$^c$, and
O. Yushchenko$^d$
\\ {\it $^a$Brookhaven National Laboratory, Upton, NY  11973}\\
$^b$ {\it International Institute of Theoretical and Applied
Physics, Ames, IA.~~50011}\\
$^c$ 
{\it Iowa State University,  Ames, IA 50011} \\
$^d$ {\it Institute for High Energy Physics,  Protvino, 142284,
Russia}}

\maketitle
\thispagestyle{empty}\pagestyle{empty}
\begin{abstract} 
We explore the capability of a $500$ or $1000$
GeV $e^+e^-$ linear collider to measure anomalous quartic gauge
boson couplings. In the framework
of a non-linear effective Lagrangian with a custodial
$SU(2)$ symmetry, there are only two 
next-to-leading order operators which
contribute to quartic, but not to two- and three- gauge
boson interactions.    The limits on the coefficients of
these operators from  present and future
$e^+e^-$ colliders  are compared with those 
available from other sources.
\end{abstract}

\section{Introduction} 
The non-Abelian structure of the Standard Model of electroweak
interactions gives precise predictions for the three- and four- gauge
boson self-interactions.  Precision tests of these interactions can
thus be used to search for new physics beyond the Standard Model. 
There exist numerous studies of the capabilities of various
machines such as the Tevatron and LEP II to measure anomalous 
three-gauge boson couplings \cite{reviews}.    However,
only very indirect limits exist on the four-gauge boson couplings
and the prospects for direct measurements have not been explored 
thoroughly in the literature. 

The direct study of four-gauge boson couplings requires the production 
of at least three gauge bosons or the observation of vector-boson 
scattering processes. For this reason they are harder to probe directly 
than their three-gauge boson counterparts 
and 
 higher energy machines are needed.
 The sensitivity of the four-gauge boson 
couplings to new physics is illustrated by the fact that the lowest 
order couplings in the standard model (without the Higgs boson) 
lead to  gauge-boson scattering amplitudes that violate unitarity 
at an energy scale around $1.8~TeV$ \cite{lqt}. In the standard model 
this bad high energy behavior is corrected by the exchange of a Higgs boson. 
In this way, gauge-boson scattering amplitudes constitute a real probe of 
the symmetry breaking sector of the model. Thus  we 
expect to learn something about the mechanism of electroweak symmetry 
breaking by studying the four gauge boson couplings. 

Here,  we consider three-gauge boson production in $e^+e^-$ 
collisions in the processes, $e^+e^- \rightarrow ZZZ$ and $e^+e^-
\rightarrow W^+ W^- Z$.  Three $Z$ production has a very small 
cross section in the Standard Model, making it an exceptionally
good place to search for new physics.  
The rate for $W^+W^-Z$ production  has a different dependance
on the quartic couplings and so 
by combining results from the two reactions,
significant limits on the quartic coupling constants can
be obtained.    
These processes have previously been considered by
Belanger and Boudjema \cite{bb}. We extend their 
analysis by carefully considering the details of the two reactions. 
In particular we study the possibility of improving the bounds that 
can be placed on quartic couplings by imposing different kinematic cuts.   
We find that unlike the case 
of triple-gauge boson couplings, it will be very difficult to 
improve the sensitivity to quartic couplings by means other than 
increasing the energy of the machine.

\section{Preliminaries}  
We consider a picture in which
there is no Higgs boson at low energy and
  the new physics responsible for the
electroweak symmetry breaking occurs  at some high scale, $\Lambda 
<  
4 \pi v \sim  3~TeV$.
In this case, the physics at low energy can be written in terms of an
effective Lagrangian
describing the interactions of 
the $SU(2)_L\times U(1)_Y$ gauge fields with the
Goldstone bosons which become the longitudinal components of the
$W^\pm$ and $Z$ gauge bosons, $\omega^\pm$ and $z$.
 The minimal 
Lagrangian which describes the interactions of the $SU(2)_L\times U(1)_Y$
gauge bosons with the Goldstone bosons  is, 
\beq
{\cal L}_2={v^2\over 4}Tr\biggl[D^\mu\Sigma^\dagger D_\mu 
\Sigma\biggr]+{\rm  Kinetic~ Energy~ Terms}
\eeq
 and has been discussed in detail in Refs. \cite{longo, dv}. 
  This non-renormalizable
Lagrangian is  to be
interpreted as an effective field theory, valid below
the  scale $\Lambda$  
and  yields the gauge boson self 
interactions which  we use in this calculation \cite{dv}. 
These interactions are identical to those of the Standard
Model when the Higgs mass is taken to be very large.  

In this scenario, the effects of new physics are
described in terms of a derivative expansion in powers of $s/\Lambda^2$.
At ${\cal O}(s/\Lambda^2)$ there are 13 possible new interactions.
We will assume a custodial $SU(2)_C$ symmetry along with CP
conservation which restricts the number of operators
at this order to 5.   With these assumptions there
are only two new interactions which contribute to
four- gauge boson vertices, but not to 
two- and three- gauge boson
vertices,
\beq
{\cal L}_4={v^2\over \Lambda^2}\biggl\{
L_1Tr\biggl(D_\mu\Sigma D^\mu \Sigma^\dagger\biggr)^2
+L_2 \biggl(D_\mu\Sigma D^\nu \Sigma^\dagger\biggr)^2
\biggr\}
\quad .
\eeq 
It is these truly quartic interactions which we study.
With the normalization we have chosen, $L_1$ and $L_2$ are
naturally of ${\cal O}(1)$.  
The Feynman rules for the four- gauge boson vertices resulting
from this Lagrangian are given in the appendix of Ref. \cite{dv}.  
  
At present, the only experimental limits
on $L_1$ and $L_2$ are indirect limits from precision measurements
at LEP.  These limits arise from loop effects and depend 
logarithmically on a cutoff, which we take to be $1~TeV$.  Implicit in the 
limits obtained from the
LEP results is the assumption that there are
 no cancellations between
the effects of
different operators.  Assuming only $L_1$ and $L_2$ are non-zero, 
the $95\%$ confidence level bounds are,\footnote{We have updated
the limits of Ref. \cite{dv}.}   
\beq
-28\biggl({\Lambda\over 2~TeV}\biggr)^2
< L_1 + {5\over 2} L_2 <
19 \biggl({\Lambda\over 2~TeV}\biggr)^2
\quad .  
\label{lep}  
\eeq
  
The LHC will also be able to limit 
four- gauge boson interactions through
vector boson scattering, $p p \rightarrow V V X$.\cite{quart,bdv}
  Since the three-
gauge boson interactions also contribute to these processes, it is again
necessary to assume that there are no cancellations between the 
contributions of the different operators.  Bounding
one  coefficient
at a time, Ref. \cite{bdv} found that the LHC will be able to obtain
limits of roughly, 
\beq
L_1,L_2 <{\cal O}(1) \biggl({\Lambda\over 2~TeV}\biggr)^2
\quad .
\eeq 

\section{{\lowercase{$e^+e^-$}} $\rightarrow ZZZ$} 
In an $e^+e^-$ collider, there is no $4$ $Z$
coupling and  $3$ $Z$ production  can be computed
 at lowest order in the energy
expansion using the Lagrangian
of Eq. 1. (This is equivalent to the Standard Model with
the Higgs graphs removed or the Higgs boson taken very massive.)
We find \footnote{These numbers agree with Ref. \cite{bh}
for $M_H=1~TeV$.} 
\beq
\sigma_{\rm SM}(e^+e^-\rightarrow ZZZ)=
\left\{
\begin{array}{ll}
.8~fb & {\rm at}~\sqrt{s}=500~GeV \\
.7~fb & {\rm at}~\sqrt{s}=1~TeV
\quad .  
\end{array}
\right.
\eeq
 For an integrated luminosity of $50~fb^{-1}$ at $\sqrt{s}=500~GeV$,
this yields  only $400$ events. 
When we  include final state branching
ratios and detector
efficiencies,  it is apparent
 that this process will be statistics limited.  

To explore new
physics beyond the lowest order in the energy 
expansion, we include the interactions of Eq. 2.  It is
important to note that there are no contributions from $3$ gauge
boson vertices to this process at this order in the energy
expansion, so 
$e^+e^-\rightarrow ZZZ$  is an unambiguous test of
the quartic $ZZZZ$ vertex  
and   gives 
 limits  on
the couplings $L_1$ and $L_2$ without the
assumption of naturalness. 
It is this feature which makes the $3Z$ production so useful for 
exploring new physics.\footnote{The $e^+e^-Z$ vertex is renormalized
by a factor sensitive to the coupling $L_{10}$.\cite{dv}.  However,
$L_{10}$ is severely restricted by LEP measurements (since it
contributes to the gauge boson two-point functions) and so
the inclusion of this operator would not change our results.}   

We include in our calculation
  an error on the luminosity measurement of $.1\%$ and a
realistic efficiency for  the $ZZZ$ final state reconstruction of
\beq
\epsilon_{ZZZ}=15\%
\eeq
as estimated for anticipated NLC detectors \cite{eff1}.  
\begin{figure}[b]
\epsfig{file=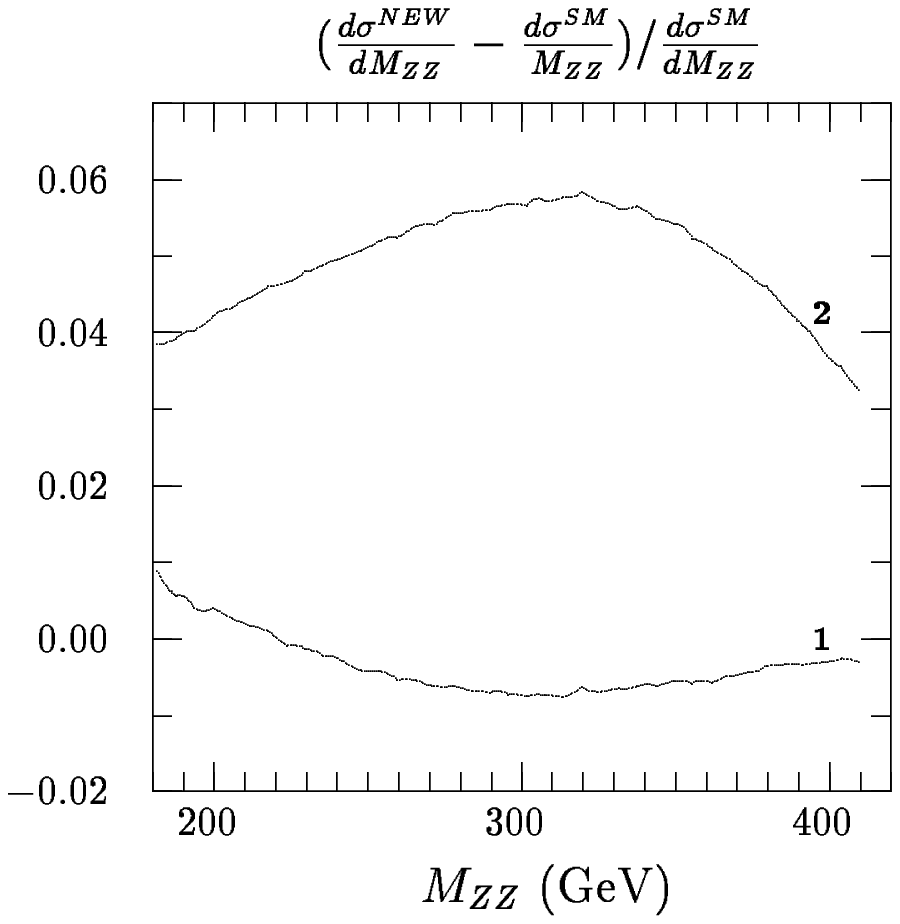,height=3.5in}
\caption{$d\sigma/dM_{ZZ}$ for $e^+e^-\rightarrow
ZZZ$ at $\protect\sqrt{s}=
500~GeV$.  The line labelled $1$ ($2$) has
$L_1+L_2=3~(-3)$. This figure assumes $\Lambda=2~TeV$.
$\sigma^{\rm {NEW}}$ is the cross section with the
interactions of Eqs. 1 and 2, while $\sigma^{\rm {SM}}$
includes only the interactions  of Eq. 1. }
\end{figure}  
In Fig. 1, we demonstrate the effects of non-zero $L_1$ and
$L_2$ and plot  $d\sigma/dM_{ZZ}$ (where $M_{ZZ}$
is the invariant mass of any pair of $Z$'s).
  We see that the effects
are rather small, of order a few percent.  This is  unfortunately
 the same
with all the distributions we have considered.
We also tried to implement some kinematic cuts to
enhance the sensitivity to $L_1$ and $L_2$.  The
most effective values of the cuts can be obtained by
studying the statistical significance which is defined as,
\beq
S\equiv  {\mid \sigma^{\rm {NEW}}-\sigma^{\rm {SM}}
\mid  
\over \sqrt{\sigma^{\rm {SM}}}} \sqrt{ { \cal L}}
\quad ,
\eeq
where ${\cal L}$ is the integrated luminosity.  
\begin{figure}[b]
\epsfig{file=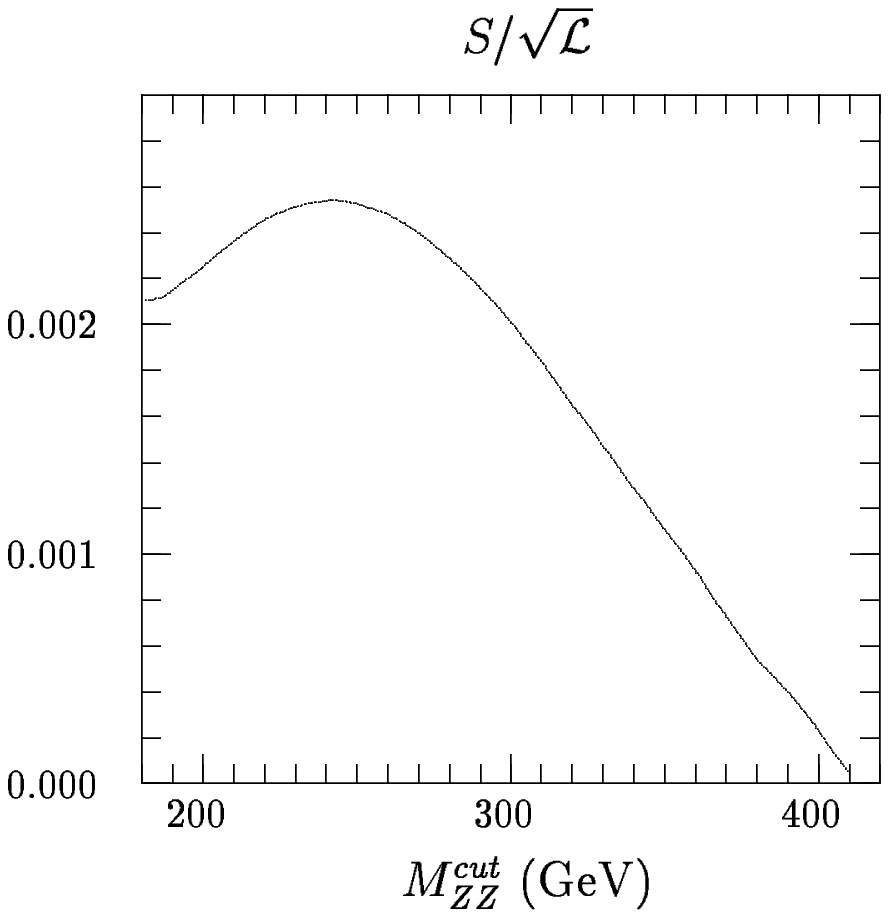,height=3.5in}
\caption{Statistical significance for the signal
using a cut on $M_{ZZ}$ at $\protect\sqrt{s}=500~GeV$
 from $e^+e^-\rightarrow
ZZZ$.  This figure assumes $L_1+L_2=3$ and $\Lambda=2~TeV$.}
\end{figure}  
In Fig. 2 we show the dependance of the statistical significance
$S$ on the chosen $M_{ZZ}$ cut.  The presence of the clear peak
at $M_{ZZ}^{cut}\simeq 240~GeV$ implies that this is the optimal
 value of the cut  for enhancing the sensitivity to the new physics effects.
\begin{figure}[b]
\epsfig{file=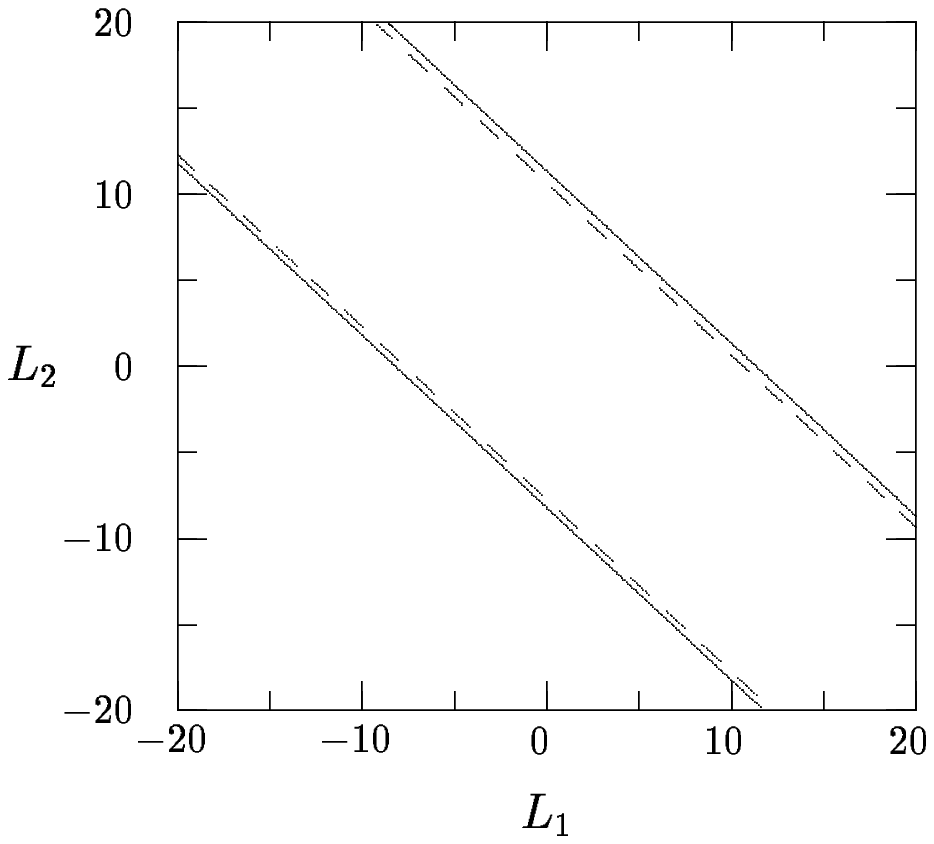,height=3.5in}
\caption{$95\%$ confidence level bounds on $L_1$ and $L_2$
from $e^+e^-\rightarrow ZZZ$ at 
$\protect\sqrt{s}=500~GeV$ .  The solid (dashed)
lines show the bound with (without) the cut on $M_{ZZ}$.
This figure assumes $\Lambda=2~TeV$. }
\end{figure}  
In Fig. 3, we  show
the allowed regions (at $95\%$ confidence level)
 of $L_1$ and
$L_2$ which are obtainable with  and without 
 the cut on $M_{ZZ}$.  We see that even though Fig. 2
demonstrates the usefulness of the kinematic cut, the 
bound of Fig. 3 which results from utilizing
 the entire kinematic region is 
more stringent than that obtained by imposing
the cut.  This is because imposing kinematic  cuts leads
to a reduction in the cross section and consequently to an increase
in the statistical error.  Hence, with such a small cross section
as in $3$ $Z$ production, the cuts are ineffective
for all distributions which we have considered.  
 
 The bounds at $\sqrt{s}=
1~TeV$ are considerably more stringent than
those at lower energy.\footnote{
We have taken $ {\cal L}=50~fb^{-1},~100~fb^{-1}$ for 
$\sqrt{s}=500~GeV,~1~TeV$.}
 The region  which can be probed
depends linearly on $L_1+L_2$ and 
for $\sqrt{s}=500~GeV$, we obtain roughly,
\beq
\mid L_1+L_2\mid < 10 \biggl({\Lambda\over 2~TeV}\biggr)^2
\quad .
\eeq 
This is only a slight improvement over the LEP limits of 
Eq.~\ref{lep}, but 
is a factor of $4$ improvement of the limits obtained
in Ref. \cite{bb}.

The dependance on $L_1+L_2$
 can easily be understood by computing the rate for $e^+e^-
\rightarrow Z zz$ using  the electroweak equivalence theorem.
At high energy, this is the dominant process contributing to
$3~Z$ production.  
In this case, there is a single
 $s$-channel diagram for $e^+e^-\rightarrow
Z^*\rightarrow Z zz$  and the 
$ZZzz$ vertex depends only on the combination   $L_1+L_2$.
From Ref. \cite{bdv}, the amplitude for
$Z^{\mu}(q_1)Z^{\nu}(q_2)\rightarrow z(p)z(p^\prime)$ is
\beqn
{\cal A}(ZZ\rightarrow zz)&=& {-g^2\over 2 \pi^2 v^2}
{L_1+L_2\over \cos^2\theta_W}\biggl[
{s\over 2} g^{\mu\nu}+p^\mu p^{\prime \nu}
+p^{\nu} p^{\prime\mu}\biggr]
\nonumber \\
&&+{g^2\over \pi^2 v^2 \cos^2\theta_W}
\biggl[ -{s\over 2} g^{\mu\nu} +q_2^{\mu}q_1^{\nu}\biggr]
\nonumber \\
&&
\cdot  
(1-4 \sin^2\theta_W\cos^2\theta_W)
\nonumber \\
&&+{g^2\over 16\pi^2 v^2}
\biggl[
g^{\mu\nu}{s\over 2}\log\biggl({s\over\mu^2}\biggr)
\nonumber \\&&  
+p^\nu p^{\prime \mu}
\log\biggl({-u\over \mu^2}\biggr)
+p^\mu p^{\prime\nu}\log
\biggl({-t\over \mu^2}\biggr)\biggr]
, 
\nonumber \\  && 
\label{zzzz}
\eeqn 
where $s=2 p\cdot p^\prime$,
$t=(q_1-p)^2$,
$u=(q_1-p^\prime)^2$, and $\mu$ is
an arbitrary renormalization scale.  
It is straightforward to turn this amplitude into a
cross section for $e^+e^-\rightarrow Z zz$ and
the results yield good agreement with those shown in
Fig.~3.  The first
line in Eq.~\ref{zzzz}
  is the contribution from the ${\cal O}({s\over \Lambda^2})$
Lagrangian of Eq. 2, while the others  are the one-
loop corrections obtained using the lowest order Lagrangian of
Eq. 1.  Both sets of terms are formally of the same order in the
energy expansion and must be included for a consistent
analysis.  
 Numerically we find, however, that for the
range of values of $L_1+L_2$ being probed at possible next linear
colliders, the loop corrections to the amplitude are extremely small
when compared to the contribution from the $L_1+L_2$ term.
This gives us confidence that the results shown in Fig. 3 will not
be significantly altered by the inclusion of electroweak radiative
corrections in the analysis. 
\begin{figure}[b]
\epsfig{file=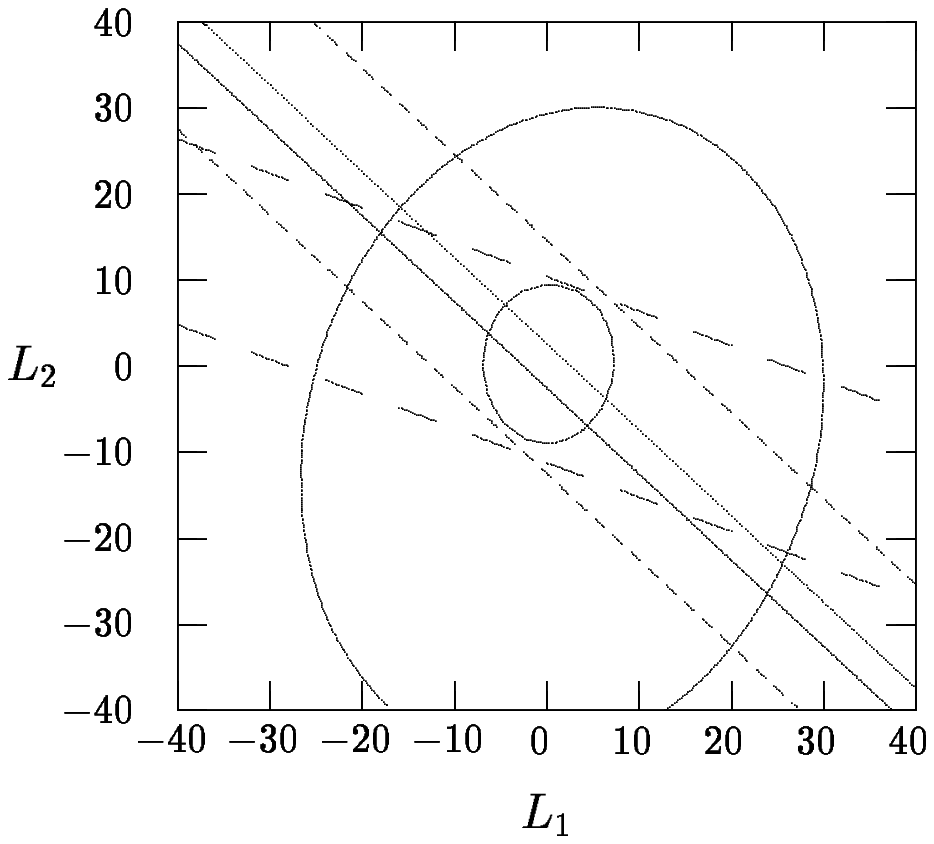,height=3.5in}
\caption{$95\%~ CL$ bounds on $L_1$ and $L_2$.  The short-
dashed (solid) bands are from $e^+e^- \rightarrow ZZZ$
at $\protect\sqrt{s}=
500~GeV$ ($1~TeV$) and the large (small)
ellipse are from $e^+e^- \rightarrow W^+W^-Z$ at
$\protect\sqrt{s}=500~GeV$ ($1~TeV$). The long dashed band is
the limit from LEPI.
This figure assumes $\Lambda=2~TeV$. 
} 
\end{figure}  
  
\section{{\lowercase{$e^+e^-$}}$\rightarrow W^+W^-Z$} 
  
The process $e^+e^-\rightarrow W^+W^-Z$ has a larger cross section
than that for $e^+e^-\rightarrow ZZZ$,\footnote{These
results agree with Ref. \cite{bh} for $M_H=1~TeV$.}   
\beq
\sigma(e^+e^-
\rightarrow W^+W^-Z)_{SM}=
\left\{ \begin{array}{ll}
43~fb & {\rm at} ~\sqrt{s}=500~GeV\\
70~fb  & {\rm at} ~\sqrt{s}=1~TeV
\quad .
\end{array}
\right.  
\eeq  
  We see that
the production of $W^+W^-Z$ depends on the three- gauge 
boson couplings as well as the quartic couplings.
Hence to obtain limits on the quartic
couplings  from this process
requires that we assume that there are  no
cancellations between the various terms, an
assumption which was not necessary in the $3~Z$ case.  We
assume that the three- gauge boson vertices have their Standard
Model forms and consider only the effects of non-zero $L_1$ and 
$L_2$.

To model detector efficiencies, we again  use an efficiency for
the $WWZ$ reconstruction  modeled on potential NLC detectors
\cite{eff}
and take 
\beq
\epsilon_{WWZ}=12\%
\quad .  
\eeq
As in the process $e^+e^-\rightarrow ZZZ$,
 the kinematic cuts are ineffective
for extracting the effects of $L_1$ and $L_2$, which are 
again rather small.  Some insight can be
gained by  writing  the cross section (in $fb$) at
$\sqrt{s}=500~GeV$ as,
\vspace*{.4in}  
\beqn 
\sigma(e^+e^-\rightarrow W^+W^-Z)&=& 
43-.022L_1+.049L_2
\nonumber \\ 
&&
 -.001L_1L_2
 +.006L_1^2+.003 L_2^2
\nonumber \\
&&
\eeqn  
The smallness of the coefficients of the $L_i$
makes it apparent that the effects of the new physics
are  extremely small.

  In Fig. 4, 
we show the $95\%$ confidence level bounds  on $L_1$ and $L_2$ 
resulting from combining the
$ZZZ$ and the $W^+W^-Z$ process. 
We also compare with the bounds previously obtained from LEPI.

\section{Conclusions}
By combining the results from $e^+e^-\rightarrow ZZZ$ and
$e^+e^-\rightarrow W^+W^-Z$, we see that an NLC will be
sensitive to a small region in the $L_1$, $L_2$ plane.  
This limit will be a considerable improvement over the
present indirect limit from precision measurements at LEP.  
In addition, 
the limit from $e^+e^-\rightarrow ZZZ$ is cleaner theoretically
than  other limits since there is no dependance on
three- gauge boson couplings.

\end{document}